\documentstyle[12pt]{article}

\newcommand{\beq}[1]{\begin{equation}\label{#1}}
\newcommand{\eeq}{\end{equation}}
\newcommand{\bear}[1]{\begin{eqnarray}\label{#1}}
\newcommand{\ear}{\end{eqnarray}}
\newcommand{\nn}{\nonumber}

\catcode`\@=11 \@addtoreset{equation}{section}\catcode`\@=12
\newcommand{\be}{\begin{equation}}
\newcommand{\ee}{\end{equation}}
\newcommand{\ba}{\begin{eqnarray}}
\newcommand{\ea}{\end{eqnarray}}

\newcommand{\np}{ {\newpage } }

\newcommand{\N}{ \mbox{\rm I$\!$N} }
\newcommand{\R}{ \mbox{\rm I$\!$R} }
\def\C{\mbox{\rm {I\kern-.520em C}}}

\newcommand{\eps}{ \varepsilon }

\newcommand{\p}{\partial}
\newcommand{\btd}{\bigtriangledown}
\newcommand{\btu}{\bigtriangleup}

\begin{document}

\centerline{\large \bf Composite p-branes on Product of Einstein Spaces}

\vspace{1.03truecm}

\bigskip

\centerline{\large 
V. D. Ivashchuk }

\vspace{0.96truecm}

\centerline{Center for Gravitation and Fundamental Metrology}
\centerline{VNIIMS, 3-1 M. Ulyanovoy Str.}
\centerline{Moscow, 117313, Russia}
\centerline{e-mail: ivas@rgs.phys.msu.su}

\begin{abstract}

A multidimensional gravitational model with
several scalar fields, fields of forms and cosmological
constant is considered. When scalar fields are  constant
and   composite $p$-brane monopole-like
ansatz for the fields of forms is adopted,
a wide class of solutions on
product of $n+1$ Einstein spaces is obtained. These solutions
are composite $p$-brane generalizations of 
the  Freund-Rubin solution.
Some examples including the $AdS_{m}  \times S^k \times \ldots$ solutions 
are considered.

\end{abstract}

\np

\section{\bf Introduction}
\setcounter{equation}{0}

Recently an interest to  Freund-Rubin type solutions \cite{FR,E,DNP}
in multidimensional models with $p$-branes ``living''
on product of Einstein spaces appeared (see, for example, 
\cite{torin,DLP,BPS} and references therein). This interest was 
inspired by papers devoted to duality between a certain limit of some
superconformal theory in $d$-dimensional space and 
string or M-theory compactified on the space $AdS_{d+1} \times
W$, where $AdS_{d+1}$ is $(d+1)$-dimensional anti-de Sitter space
and $W$ is a compact manifold (e.g. sphere $S^m$) \cite{M}
(see also \cite{FF,CKKTP,GKP,W,AV1,AOY} etc.)

In this paper we obtain a rather general class of solutions 
defined on product of $n +1$ Einstein spaces  for the multidimensional
gravitational model with fields of forms and scalar fields. These
solutions generalize the  Freund-Rubin solutions \cite{FR} to
the composite $p$-brane case with constant scalar fields. They follow
just from the equations of motion when certain restriction on
intersections of ``extended'' $p$-brane worldvolumes is imposed.
In the non-composite case
the ``cosmological derivation'' of some special 
(static) solutions with $p$-branes
was performed  in \cite{Cosm}. 
 
We note that in the pure gravitational model with cosmological 
constant the solutions describing the product of
Einstein spaces were considered in 
\cite{BIMZ,Zh}. These solutions were also generalized to
some other matter fields, e.g. scalar one (see \cite{GZ}
and references therein).

It was shown in \cite{GT} that the solutions in $D = 10,11$ supergravities
representing $D3, M2, M5$ branes interpolate between flat-space vacuum and
compactifications to AdS space. The AdS spaces appear in the ``near-horizon''
limit. 
The solutions obtained here are also related to the so-called
Madjumdar-Papapetrou type solutions with intersecting
composite p-branes (see \cite{AR,IM3,IMBl} and references therein).
This correspondence 
will be considered in detail in a separate publication \cite{IMF}. 

We note that in \cite{IMBl} a large variety of so-called
``block-orthogonal'' Madjumdar-Papapetrou (MP) type
$p$-brane solutions was obtained. These solutions may be
related to Lie algebras: simple or hyperbolic \cite{IMBl,IKM}.
The non-extremal "block-orthogonal" solutions were obtained
earlier in \cite{Br}. The ``block-orthogonal'' MP solutions
contain the ``orthogonal'' ones (see \cite{PT,Ts,GKT,S,G,AV2,IM2,AR,IM3})
and references therein) as a special case. 

The Freund-Rubin solutions \cite{FR}  in $D =11$
supergravity \cite{CJS}: $AdS_{4} \times S^7$ and  $AdS_{7} \times S^4$,   
correspond to ``electric'' $M2$-branes
``living'' in  $AdS_4$ and $S^4$ respectively or, equivalently,
to ``magnetic'' $M5$ branes ``living'' in $S^7$ and $AdS_7$ respectively).
The ``popular'' $AdS_{5} \times S^5$ solution in 
$II B$ ($D =10$) supergravity model ( see, for example \cite{DLP,BPS})
corresponds to a composite self-dual configuration
with two $D3$ branes living in $AdS_{5}$ and $S^5$ respectively
and corresponding to a 5-form.

In Sect. 2 we outline the general approach with arbitrary forms
and dilaton fields on a product of $(n+1)$ manifolds. 
In Sect. 3 we give a general solution for static fields on
a product of $(n+1)$ Einstein spaces.  
Several examples of solutions are presented. Among
them the solution of $D = 11$ supergravity on the manifold
$AdS_{2}  \times S^2 \times M_{2}  \times M_{3}$ is considered.
($AdS_{k}  \times S^l \times T^{m}$
solutions  of $D =11$ supergravity and others were 
listed in \cite{DLP,BPS}).
This solution corresponds to the ``near horizon'' limit  of
the  Madjumdar-Papapetrou type solution
of $D = 11$ supergravity describing  a bound state of $M2$ and 
$M5$ branes  \cite{IMBl} with the 
intersection rule corresponding to the Lie algebra $A_2 = sl(3)$ \cite{IMJ}.

\section{\bf The model}
\setcounter{equation}{0}

We consider the model governed by the action
\bear{2.1}
S =&& \int_{M} d^{D}z \sqrt{|g|} \{ {R}[g] - 2 \Lambda - h_{\alpha\beta}
g^{MN} \partial_{M} \varphi^\alpha \partial_{N} \varphi^\beta
\\ \nn
&& - \sum_{a \in \Delta}
\frac{\theta_a}{n_a!} \exp[ 2 \lambda_{a} (\varphi) ] (F^a)^2 \},
\ear
where $g = g_{MN} dz^{M} \otimes dz^{N}$ is the metric,
$\varphi=(\varphi^\alpha) \in \R^l$
is a vector from dilatonic scalar fields,
$(h_{\alpha\beta})$ is a non-degenerate 
symmetric $l\times l$ matrix ($l\in \N$), $\theta_a \neq 0$, 
\beq{2.2} 
F^a =  dA^a =\frac{1}{n_a!} F^a_{M_1 
\ldots M_{n_a}} dz^{M_1} \wedge \ldots \wedge dz^{M_{n_a}} 
\eeq
is a  $n_a$-form ($n_a \geq 2$) on a $D$-dimensional manifold $M$, 
$\Lambda$ is 
a cosmological constant and $\lambda_{a}$ is a $1$-form on $\R^l$ :  
$\lambda_{a} (\varphi) =\lambda_{a \alpha}\varphi^\alpha$,
$a \in \Delta$; $\alpha=1,\ldots,l$.
In (\ref{2.1})
we denote $|g| = |\det (g_{MN})|$,
\beq{2.3}
(F^a)^2 =
F^a_{M_1 \ldots M_{n_a}} F^a_{N_1 \ldots N_{n_a}}
g^{M_1 N_1} \ldots g^{M_{n_a} N_{n_a}},
\eeq
$a \in \Delta$, where $\Delta$ is some finite set.
In the models with one time all $\theta_a =  1$  
when the signature of the metric is $(-1,+1, \ldots, +1)$.

The equations of motion corresponding to  (\ref{2.1}) have the following
form
\bear{2.4}
R_{MN}  =   Z_{MN} + \frac{2\Lambda}{D - 2} g_{MN},
\\
\label{2.5}
{\btu}[g] \varphi^\alpha -
\sum_{a \in \Delta} \theta_a  \frac{\lambda^{\alpha}_a}{n_a!}
e^{2 \lambda_{a}(\varphi)} (F^a)^2 = 0,
\\
\label{2.6}
\nabla_{M_1}[g] (e^{2 \lambda_{a}(\varphi)}
F^{a, M_1 \ldots M_{n_a}})  =  0,
\ear
$a \in \Delta$; $\alpha=1,\ldots,l$.
In (\ref{2.5}) $\lambda^{\alpha}_{a} = h^{\alpha \beta}
\lambda_{\beta a}$, where $(h^{\alpha \beta})$
is a matrix inverse to $(h_{\alpha \beta})$.
In (\ref{2.4})
\beq{2.7a}
Z_{MN}= Z_{MN}[\varphi] +  
\sum_{a \in \Delta} \theta_a e^{2 \lambda_{a}(\varphi)} Z_{MN}[F^a,g],
\eeq
where
\bear{2.7}
Z_{MN}[\varphi] =
h_{\alpha\beta} \p_{M} \varphi^{\alpha} \p_{N} \varphi^{\beta},
\\  \label{2.8}
Z_{MN}[F^a,g] = \frac{1}{n_{a}!}  \left[ \frac{n_a -1}{2 -D}
g_{MN} (F^{a})^{2}
 + n_{a}  F^{a}_{M M_2 \ldots M_{n_a}} F_{N}^{a, M_2 \ldots M_{n_a}}
 \right].
\ear

In (\ref{2.5}) and (\ref{2.6}) ${\btu}[g]$ and ${\btd}[g]$
are Laplace-Beltrami and covariant derivative operators respectively
corresponding to  $g$.

{\bf  Multi-index notations.} Let us consider the manifold  
\beq{1.1}
M = M_{0}  \times M_{1} \times \ldots \times M_{n}.
\eeq
We denote $d_{i} = {\rm dim} M_{i} \geq 1$; $i = 0, \ldots, n$.
$D =  \sum_{i = 0}^{n} d_{i}$.
Let $g^i  = g^{i}_{m_{i} n_{i}}(y_i) dy_i^{m_{i}} \otimes dy_i^{n_{i}}$
be a metric on the manifold $M_{i}$, $i=0,\ldots,n$.
Here we use the notations of our previous papers \cite{IM1,IM2,IMR,IM3}.
Let any manifold $M_{\nu}$  be oriented and connected.
Then the volume $d_i$-form
\beq{2.A}
\tau_i  \equiv \sqrt{|g^i(y_i)|}
\ dy_i^{1} \wedge \ldots \wedge dy_i^{d_i},
\eeq
and the signature parameter
\beq{1.5}
\varepsilon_{i}  \equiv {\rm sign}( \det (g^i_{m_i n_i})) = \pm 1
\eeq
are correctly defined for all $i = 0,\ldots,n$.

Let $\Omega = \Omega(n+1)$  be a set of all non-empty  
subsets of $\{ 0, \ldots,n \}$.
The number of elements in $\Omega$ is $|\Omega| = 2^{n + 1} - 1$.
For any $I = \{ i_1, \ldots, i_k \} \in \Omega$, $i_1 < \ldots < i_k$,
we denote
\bear{1.6}
\tau(I) \equiv \hat{\tau}_{i_1}  \wedge \ldots \wedge \hat{\tau}_{i_k},
 \\ \label{1.6a}
\eps(I) \equiv \eps_{i_1}  \times  \ldots \times \eps_{i_k},
 \\ \label{1.7}
M_{I} \equiv M_{i_1}  \times  \ldots \times M_{i_k},
\\ \label{l.8}
d(I) \equiv  \sum_{i \in I} d_i,
\ear
where $d_i$ is both, the dimension of the oriented manifold $M_i$
and the rank of the volume form $\tau_i$ and $\hat{\tau}_i$
is the pullback of $\tau_i$ to the manifold $M$:
$\hat{\tau}_i = p_i^{*} \tau_iÿ$, where
$p_{i} : M \rightarrow  M_{i}$, is the canonical projection,
$i = 0, \ldots, n$.

We also denote by
\beq{1.9}
\delta_{I}^i= \sum_{j\in I} \delta_{j}^i
\eeq
the indicator of $i$ belonging
to $I$: $\delta_{I}^i =1$ for $i\in I$ and $\delta^{i}_{I}=0$ otherwise.

\section{The solution}

The solution reads as following. The metric
is defined on the manifold (\ref{1.1}) and has the following form
\beq{1.2}
g= \hat{g}^0  + \hat{g}^1 + \ldots + \hat{g}^n,
\eeq
where  $g^i$  is a metric on $M_{i}$  satisfying the equation
\beq{1.3}
{\rm Ric}[g^i]= \xi_{i} g^i,
\eeq
$\xi_{i} = {\rm const} $,
$i=0,\ldots,n$. Here ${\rm Ric}[g^{i}]$ is Ricci-tensor corresponding 
to $g^{i}$ and $\hat{g}^{i} = p_{i}^{*} g^{i}$ is the 
pullback of the metric $g^{i}$  to the manifold  $M$ by the 
canonical projection: $p_{i} : M \rightarrow  M_{i}$, $i = 0, 
\ldots, n$. Thus, all $(M_{i}, g^{i})$  are Einstein spaces.

The fields of forms and  scalar fields are the following
\bear{2.9}
F^a = \sum_{I \in \Omega_{a}} Q_{aI} \tau(I),  
\\ \label{2.10}
\varphi^{\alpha} = {\rm const}.
\ear
where $Q_{aI}$ are constants,
$\Omega_{a} \subset \Omega$ are  subsets, 
satisfying the relations
\beq{2.23}
d(I) = n_a ,
\eeq
$I \in \Omega_a$, $a \in \Delta$, and
the {\bf Restriction} presented below.
The  parameters of the solution obey the relations
\bear{2.11}
\sum_{a \in \Delta} \theta_a \lambda^{\alpha}_a
e^{2 \lambda_{a}(\varphi)} 
\sum_{I \in \Omega_a} (Q_{aI})^2 \eps(I) = 0, \\
\label{2.12}
\xi_i =   \frac{2 \Lambda}{D-2}  +
\sum_{a \in \Delta} \theta_a e^{2 \lambda_{a}(\varphi)} 
\sum_{I \in \Omega_a} (Q_{aI})^{2} \eps(I) \big[ \delta^i_I
- \frac{n_a -1}{D - 2} \big],
\ear
$i = 0, \ldots, n$.

The solution is valid if the following restriction
on the sets $\Omega_{a}$, $a \in \Delta$, similar to that
from \cite{IM3} (see also \cite{AR}) is satisfied.

{\bf Restriction.} 
For any  $a \in \Delta$ and $I,J \in \Omega_{a}$, $I \neq J$, we put
\beq{2.r} 
d(I \cap  J) \leq n_a - 2.
\eeq

This restriction guarantees the block-diagonal structure
of the  $Z_{MN}$-tensor in (\ref{2.7a}) (see  
relation (\ref{A.6}) from the Appendix).

The solution mentioned above may be verified by a straightforward
substitution of the fields from (\ref{1.2})-(\ref{2.10})
into equations of motion (\ref{2.4})-(\ref{2.8}) while formulas
from the Appendix are keeping in mind. We note that
due to the relations $dF^a = 0$ the potential form $A^a$ satisfying
$F^a = dA^a$ exists at least locally, $a \in \Delta$.

We note that the {\bf Restriction} is satisfied
if the number of $1$-dimensional
manifolds among $M_i$ is no more than $1$.

\subsection{``Electro-magnetic'' form of solution}

Due to relation
\beq{2.24}
* \tau(I) =  \eps(I) \delta(\bar{I},I) \tau(\bar{I}),
\eeq
where $*=*[g]$ is the Hodge operator on $(M,g)$, 
\beq{2.24b}
\bar{I} 
= \{0, \ldots,n \} \setminus I 
\eeq
is ``dual''
set and $\delta(\bar{I},I) = \pm 1$  is defined by 
the following relation
\beq{2.24a}
\tau(\bar{I}) \wedge \tau(I)=  \delta(\bar{I},I) \tau(\{0, \ldots,n \}),
\eeq
the electric ``brane'' living in $M_I$ (see (\ref{1.7}))
may be interpreted also as a magnetic ``brane'' living 
in $M_{\bar{I}}$.  The relation (\ref{2.9})
may be rewritten in the ``electro-magnetic'' form as following
\beq{2.25}
F^a = \sum_{I \in \Omega_{ae}} Q_{aIe} \tau(I)
+ \sum_{J \in \Omega_{am}} Q_{aJm} * \tau(J), 
\eeq
where   $\Omega_{a} = \Omega_{ae} \cup  \bar{\Omega}_{am}$, 
$ \Omega_{ae} \cap \bar{\Omega}_{am} = \emptyset$, 
$\bar{\Omega}_{am} \equiv \{J| J = \bar{I}, I \in \Omega_{am}  \}$,
and $Q_{aIe} = Q_{aI}$ for $I \in \Omega_{ae}$   and
$Q_{aJm} = Q_{a \bar{J}} \eps(J) \delta(\bar{J},J)$ for $J \in \Omega_{am}$.

\section{Some examples}

Here we consider some examples of the obtained solutions
when $\eps_0 = -1$ and all $\eps_i = 1$,  
$i = 1, \ldots, n$,
i.e. ``our space'' $(M_0,g^0)$ is pseudo-Euclidean space and the
``internal spaces'' $(M_i,g^i)$ are Euclidean ones. We also put
$\theta_a =  1$ and $n_a < D -1$ for all $a \in \Delta$. 

\subsection{Solution with one $p$-brane} 

Let $\Omega_a = \{ I \}$,  
$\lambda_{a} = 0$ for some  $a \in \Delta$ and
$\Omega_b$ are empty for all  $b \neq a$, $b \in \Delta$.
Equations (\ref{2.11}) are satisfied identically in this case
and (\ref{2.12}) read
\beq{2.26}
\xi_i =    \frac{2 \Lambda}{D-2}  + 
\eps(I)Q^{2} \big[ \delta^i_I - \frac{n_a -1}{D - 2} \big],
\eeq
$i = 0, \ldots, n$, where $Q = Q_{aI}$.  

\subsubsection{$p$-brane does not ``live'' in $M_0$} 

For $I = \{1, \ldots, k \}$,
$1 \leq k \leq n$, we get  $\eps(I) = 1$ and
\bear{2.27}
\xi_0 = \xi_{k+1} = \ldots =
\xi_n =    \frac{2 \Lambda}{D-2}  
- Q^{2} \frac{n_a -1}{D - 2}, \\ \label{2.28}
\xi_1 =  \ldots =  \xi_{k} = \frac{2 \Lambda}{D-2}  
+ Q^{2} \big[ 1 - \frac{n_a -1}{D - 2} \big].
\ear

For $\Lambda = 0$, $Q \neq 0$ we get
$\xi_0 = \xi_{k+1} = \ldots = \xi_n < 0$ and  
$\xi_1 =  \ldots =  \xi_{k} > 0$. These solutions
contain the solutions with the manifold
\beq{2.29}
M = AdS_{d_0} \times S^{d_1} \times \ldots \times 
S^{d_k} \times H^{d_{k+1}} \times \ldots \times M_n. 
\eeq 
Here $H^d$ is $d$-dimensional Lobachevsky space;
$M_n = H^{d_{n}}$ for $k < n$ and $M_n = S^{d_{n}}$ for $k = n$.

For $2\Lambda = Q^2 (n_a - 1)$ we get a solution with a flat 
our space: $M = \R^{d_0} \times S^{d_1} \times \ldots \times 
S^{d_k} \times \R^{d_{k+1}} \times \ldots$. We may consider 
the fine-tuning of the cosmological constant, when    
$\Lambda$ and $Q^2$ are of the Planck order but $\xi_0$ is small enough
in agreement with observational data. 

\subsubsection{$p$-brane  ``lives'' in $M_0$}

For $I = \{0, \ldots, k \}$,
$0 \leq k \leq n$, we get  $\eps(I) = - 1$ and
\bear{2.30}
\xi_{k+1} = \ldots = \xi_n =  \frac{2 \Lambda}{D-2}  
+ Q^{2} \frac{n_a -1}{D - 2}, \\ \label{2.31}
\xi_0 =  \ldots =  \xi_{k} = \frac{2 \Lambda}{D-2}  
- Q^{2} \big[ 1 - \frac{n_a -1}{D - 2} \big].
\ear

For $\Lambda = 0$, $Q \neq 0$, we get
$\xi_{k+1} = \ldots = \xi_n > 0$ and  
$\xi_0 =  \ldots =  \xi_{k} < 0$. The solutions
contain the solutions with the manifold
\beq{2.32}
M = AdS_{d_0} \times H^{d_1} \times \ldots \times 
H^{d_k} \times S^{d_{k+1}} \times \ldots \times M_n.
\eeq 
Here $M_n = S^{d_{n}}$ for $k < n$ and $M_n = H^{d_{n}}$ for $k = n$.

For $2 \Lambda = Q^2 (D -n_a - 1)$ we get a solution with a flat 
our space: $M  = \R^{d_0} \times S^{d_1} \times \ldots \times 
S^{d_k} \times \R^{d_{k+1}} \times \ldots$. We may also consider 
the fine-tuning mechanism here. 

\subsection{Solution with two $p$-branes}

\subsubsection{Composite solution on $M_0 \times M_1$}

Let $n = 1$, $d_0 = d_1 = n_a = d$,  
$\Omega_a = \{ I_0 = \{ 0 \}, I_1 = \{ 1 \} \}$,  
for some $a$ and other $\Omega_b$ are empty. 
Denote $Q_0 = Q_{a I_0}$ and $Q_1 = Q_{a I_1}$.  
For the field of form we get from (\ref{2.9})
\beq{2.33}
F^a = Q_0 \hat{\tau}_0 + Q_1 \hat{\tau}_1.
\eeq
When $\lambda_{a} \neq 0$ the equations (\ref{2.11}) are satisfied 
if and only if   $Q_0^2 = Q_1^2 = Q^2$.
Relations (\ref{2.12}) read
\bear{2.34}
\xi_0 = \frac{2 \Lambda}{D-2}  - Q^2 e^{2 \lambda_{a}(\varphi)},
 \\ \label{2.35}
\xi_1 = \frac{2 \Lambda}{D-2}  + Q^2 e^{2 \lambda_{a}(\varphi)}.
\ear

For $\Lambda = 0$ and  $Q \neq 0$ we get 
the solution defined on the manifold $M = AdS_{d} \times S^{d}$. 
For odd $d$ the form (\ref{2.33}) is self-dual (see subsection 3.1).
The solution describes a composite $p$-brane configuration containing
$ AdS_{5} \times S^{5}$  solution in $II B$ supergravity 
as a special case.  

\subsubsection{Near-horizon limit for $A_2$-dyon in $D = 11$
supergravity} 

Here we consider the extension of the Madjumdar-Papapetrou solution \cite{MP}
to $D = 11$ supergravity, describing a bound state of  two $p$-branes:
one electric ($M2$) and 
one magnetic ($M5$)  \cite{IMBl}. This solution has an unusual
intersection rule corresponding to the Lie algebra $A_2 = sl(3)$.
The solution is defined on the manifold (\ref{1.1}) with $n = 3$,
$D = 11$ and has the following form 
\bear{2.36}
g= H^2 \hat{g}^0-H^{-2}dt\otimes dt+ \hat{g}^2+ \hat{g}^3, \\
\label{2.37}
F^a=\nu_1dH^{-1}\wedge dt\wedge
\hat{\tau}_2+ \nu_2(*_0dH)\wedge \hat{\tau}_2,
\ear
where 
$H$ is the harmonic function on $(M_0,g^0)$; metrics
$g^0, g^2, g^3$ are Ricci-flat, $\eps_1=+1$,
$\nu_1^2 = \nu_2^2 =1$, ${\rm rank } F^a=4$ and 
$d_0=3$, $d_2=2$, $d_3=5$. 

Let $g^0 = dR \otimes dR + R^2 \hat{g}[S^2]$, $H = C 
+ \frac{M}{R}$, where $C$ and $M$ are constants and
$g[S^2]$ is the metric on $S^2$. For $C = 1$ the
4-dimensional section of the metric describes extremally
charged Reissner-Nordstr\"om black hole of mass $M$ in the 
region out of the horizon: $R > 0$. 

Now we put $C = 0$ and $M = 1$ (i.e. the so-called ``near-horizon''
limit is considered). We get the solution    
\bear{2.38}
g= 
\hat{g}[AdS_2] + \hat{g}[S^2] + \hat{g}^2+ \hat{g}^3, \\
\label{2.39}
F^a= \nu_1 \hat{\tau}[AdS_2] \wedge \hat{\tau}_2
+ \nu_2 \hat{\tau}[S^2] \wedge \hat{\tau}_2
\ear
defined on the manifold 
\beq{2.40}
M = AdS_{2}  \times S^2 \times M_{2}  \times M_{3}.
\eeq
Here 
$g[AdS^2] = R^{-2} [dR \otimes dR  - dt \otimes dt]$
is the metric on $AdS_2$,
$(M_i,g^i)$  are Ricci-flat,
$i = 2,3$;  $\eps_2=+1$, $d_2=2$, $d_3=5$   and
$\nu_1^2 = \nu_2^2 =1$. 
 
{\bf Remark 1.} The solutions (\ref{2.36})-(\ref{2.40})
may be generalized to the
case of so-called 
$B_D$-models in dimension $D \ge 12$ \cite{IMJ}.
In this case ${\rm rank} F^a \in \{4,\dots, D-7 \}$,  
$d_2=a-2$, $d_3=D-2-a$, and all scalar fields are zero.
In this case the solution (\ref{2.36})-(\ref{2.37})
describes $A_2$-dyon  with electric $d_1$-brane
and magnetic $d_2$-brane, corresponding to $F^a$-form and
intersecting in 1-dimensional time manifold.
We note that $B_{12}$-model  corresponds to the low-energy limit
of the $F$-theory \cite{F-th}.

{\bf Remark 2.} For $M_2 = \R^2$ and  $M_3 = \R^2 \times M_4$, 
the metrics (\ref{2.36}) and (\ref{2.38}) may be obtained
also for the solution with two M2 branes and two M5 branes
\cite{DLP,BPS}.

\section{Appendix}

Let $F_1$ and $F_2$ be forms of rank $r$ on $(M,g)$ ($M$ is a manifold and
$g$ is a metric on it). We define
\bear{A.1}
(F_1\cdot F_2)_{MN} \equiv
{(F_1)_{MM_2\dots M_r}(F_2)_N}^{M_2\cdots M_r}
\\ \label{A.2}
F_1 F_2 \equiv (F_1\cdot F_2)_M^M  =
(F_1)_{M_1M_2\dots M_r}(F_2)^{M_1M_2\cdots M_r}.
\ear

For the volume forms (\ref{1.6}) we get 
\bear{A.3}
\frac{1}{d(I)!}(\tau(I) \tau(I))=
\eps(I), \\   
\label{A.4}
\frac1{(d(I) -1)!}( \tau(I) \cdot \tau(I))_{m_in_i} =
\eps(I) \delta_{I}^i g^{i}_{m_{i} n_{i}},
\ear
where the indices $m_i, n_i$ correspond to the manifold $M_i$,
$i = 0, \ldots, n$. The symbols $\eps(I)$ and $\delta_{I}^i$
are defined in (\ref{1.6a})  and (\ref{1.9}) respectively.

Let $I,J \in \Omega$, $I \neq J$ and $d(I) = d(J)$. Then
\beq{A.5}
\tau(I) \tau(J) = 0, 
\eeq
and due to {\bf Restriction}
\beq{A.6}
(\tau(I) \cdot \tau(J))_{MN} = 0.
\eeq

\section{Conclusions}

In this paper we obtained  exact solutions 
describing  the product of $(n+1)$ Einstein spaces 
for the gravitational model with fields of forms and 
(dilatonic) scalar fields. The solutions are 
given by the relations (\ref{1.2})-(\ref{2.12})
 and may be considered
as a composite $p$-brane generalization of 
the Freund-Rubin solutions. 

These solutions 
may be used in multidimensional cosmology
as they provide a mechanism of compensation or reduction
of the cosmological constant in our space
by the use of fine-tuning of ``big'' $\Lambda$ 
(of the Planck's order) with the ``charges'' of 
$p$-branes.

Another interesting aspect is connected with
a singling out of a special subclass of 
$AdS_{m} \times S^k \times \ldots$ 
solutions originating from the Madjumdar-Papapetrou
type solutions: ``orthogonal'' and ``block-orthogonal'' \cite{IMBl} 
in the ``near-horizon'' limit. Such compactifications
are of interest for the string-  or M-theories itself
and  for the studing of the quantum phenomena near the horizon
of the black hole. Here the compactifications inheriting
the ``non-orthogonal'' intersection rules, e.g. corresponding
to different Lie algebras (simple or hyperbolic) \cite{IMBl,IKM}
should be also considered.

\begin{center}
{\bf Acknowledgments}
\end{center}

This work was supported in part
by the DFG grant  436 RUS 113/236/O(R),
by the Russian Ministry of
Science and Technology and  Russian Foundation for Basic Research,
grant 98-02-16414. The author is grateful to V.N. Melnikov
and A.I. Zhuk for useful discussions.

\end{document}